\documentclass[twoside,leqno,twocolumn]{article}  
\usepackage{ltexpprt} 
\usepackage{hyperref,url}
\usepackage{graphicx}
\usepackage{amsfonts,amsmath}

\newcommand{\bd}[1]{\mbox{\boldmath $#1$}} 
\newcommand{\ny}{\left\{}
\newcommand{\zr}{\right\}}
\newcommand{\ang}{\mbox{$\rm \AA$}}
\newcommand{\ls}{{\tiny \( \stackrel{<}{\sim}\)}}

\begin{document}

\title{Finding Young Stellar Populations in Elliptical Galaxies from
Independent Components of Optical Spectra}
\author{Ata~Kab\'{a}n\thanks{School of Computer Science, The University of Birmingham, Birmingham B15 2TT, UK. Email: A.Kaban@cs.bham.ac.uk.} \\
\and 
Louisa A.~Nolan\thanks{School of Physics and Astronomy, The University of Birmingham, Birmingham B15 2TT, UK. Email: \{lan, somak\}@star.sr.bham.ac.uk} \\
\and 
Somak Raychaudhury$^\dagger$ 
}
\date{}

\maketitle
 
%\pagenumbering{arabic}
%\setcounter{page}{1}%Leave this line commented out.

\begin{abstract}
Elliptical galaxies are believed to consist of a single population of
old stars formed together at an early epoch in the Universe, yet
recent analyses of galaxy spectra seem to indicate the presence of
significant younger populations of stars in them. The detailed
physical modelling of such populations is computationally expensive,
inhibiting the detailed analysis of the several million galaxy spectra
becoming available over the next few years. Here we present a data
mining application aimed at decomposing the spectra of elliptical
galaxies into several coeval stellar populations, without the use of
detailed physical models.  This is achieved by performing a linear
independent basis transformation that essentially decouples the
initial problem of joint processing of a set of correlated spectral
measurements into that of the independent processing of a small set of
prototypical spectra.  Two methods are investigated: (1) A fast
projection approach is derived by exploiting the correlation structure
of neighboring wavelength bins within the spectral data. (2) A
factorisation method that takes advantage of the positivity of the
spectra is also investigated.  The preliminary results show that
typical features observed in stellar population spectra of different
evolutionary histories can be convincingly disentangled by these
methods, despite the absence of input physics. The success of this
basis transformation analysis in recovering physically interpretable
representations indicates that this technique is a potentially
powerful tool for astronomical data mining.
\end{abstract}

%------------------------------------------------------------------------- 
\section{Introduction}

The optical spectrum of a galaxy is a linear superposition of the
spectra of its billions of constituent stars.  Yet, since large
populations of stars form in galaxies at definite periods of its
lifetime, and the atomic and nuclear physics of the evolution of
stellar populations, though complex, are well understood, the detailed
modelling of composite spectra of stellar populations can be used to
yield a wealth of information about the history of a galaxy from its
spectrum.

The spectrum of a star can be modelled as a function of three
parameters-- its mass, its age and its composition (since it is made
mostly of hydrogen and helium, the last parameter is characterised by
the relative abundance of other elements, and is known as the ``chemical
abundance'').  Elliptical galaxies, which account for about 20\% for
all galaxies in the Universe, are believed to consist predominantly of
a single coeval stellar population (e.g. \cite{eggen,larson,WR}), all
formed at an early epoch in the Universe. This implies that an
elliptical galaxy can be modelled as a system of stars, all of the
same age and chemical abundance, evolving together, if validated
assumptions can be made about the distribution of stellar masses.
However, as a result of detailed spectral studies conducted in the
last decade (e.g. \cite{kauff}), it now transpires that elliptical
galaxies are more complex objects, at least some of which have
undergone more recent bursts of substantial star formation, and
consequently are likely to contain more than one stellar population
component.

The determination of the star formation history of a galaxy has
important implications for the still-controversial issue of the
formation and evolution of galaxies.  Until recently, the analysis of
a large statistical sample of stellar populations of galaxies would
not have been possible since only small ensembles of galaxy spectra
were available.  However, the development of data mining tools for
automating parts of the analysis is becoming more and more essential
in the light of the rapid increase in the availability of data that is
approaching.

Recent and ongoing galaxy spectral surveys (2dFGRS,
\url{www.mso.anu.edu.au/2dFGRS/} (completed in 2003) and SDSS, 
\url{www.sdss.org/}) will produce more
than two million galaxy spectra in the next few years, which is to be
integrated into a more ambitious database of publicly-available
astronomical data, incorporating Grid technology (the {\it Virtual
Observatory}, \url{www.ivoa.net}).  Since the detailed physical
modelling of stellar populations is numerically expensive, even a
simple question like ``what fraction of elliptical galaxies contain a
significantly younger stellar population?''  will take years to
address by conventional modelling techniques using 
stellar population synthesis. 
The timely extraction of useful knowledge, such as the
characteristics of the star formation history (ages, chemical
abundances and masses of the component stellar populations) of
galaxies, from these data will largely depend on developing
appropriate data analysis tools that are able to complement more
specialised astrophysical analyses.

The astrophysical questions motivating this study are:

\begin{enumerate}
\item Can we disentangle major stellar population components of elliptical galaxies
without the use of detailed physical models?  
\item How do the results from a data-driven analysis of
observed galaxy spectra correlate with the parameters of star
formation history determined via a completely independent model
fitting technique used in astrophysics?
\end{enumerate}

These questions have not been addressed before in a data driven
manner --- that is, based on the data characteristics only,  
without any specialised physical input.  

\subsection{Roadmap}
In this paper, we discuss and investigate statistical methods
that attempt to solve the described inverse modelling problem by relating
multivariate observations to lower-dimensional vectors of
statistically independent unobserved variables through the use of a
linear model. The required statistical assumptions will be derived
from general characteristics of the data, in order to employ these
methods in an appropriate manner.

The preliminary results presented in the next sections are based on
the data described in Section~2. A projection approach that exploits
the correlation structure of the spectra is presented in Section~3. In
this approach, the required assumption for solving the inverse
modelling is derived from exploiting the correlation structure between
neighboring wavelength bins, which comes naturally with spectral
data. The independent spectral components obtained turn out to be also
physically interpretable and exhibit typical features of spectra of
the young and mature stellar populations.  We then compare the results
with a positivity-based single stage approach, presented in Section 4,
that has been often employed for analysing spectral data in different
domains \cite{juvela, Lee_Seung}. We provide a simple probabilistic
reformulation of this method that highlights its implicit assumptions,
links it to the methods developed in
\cite{Miskin} and also allows us to potentially incorporate measurement errors
(if known from domain knowledge) into the algorithm.
In Section 5, the results
are presented, discussed and comparatively assessed, first in a
data-driven manner and then, more importantly, from the astrophysical
perspective.  Finally, our conclusions are summarised in the last
section.

\section{The data and model setting}

The data we use here represent the {\it observed} optical spectra of
21 nearby elliptical galaxies, compiled by blending together 5855
measurements over the 
range 2000-8000 \AA\ from
various observatories on ground and in space. % (the UV part of the
These represent the following galaxies: NGC 0205, NGC 0224, NGC 1052, NGC 1400, NGC 1407, IC 1459, NGC 1553, 
NGC 3115, NGC 3379, NGC 3557, NGC 3605, NGC 3904, NGC 3923, NGC 4374, NGC 4472, NGC 4621, NGC 4697, NGC 5018,
NGC 5102, NGC 7144 and NGC 7252.
The spectra have been corrected for redshift (i.e. converted from their observed wavelengths to their
emitted wavelengths), and the fluxes are
normalised to unity in the region 5020-5500 \AA.

Since these spectra are compiled from sources with varying spectral
coverage, the resulting data matrix has 1453 missing values, which are
first imputed using a KNN imputation \cite{missing} from synthetic
data. We preferred this non-parametric procedure here, as the missing 
data mechanism may not be random --- an assumption
made by most of other imputation schemes. The validity of the 'missing at random' assumption 
in the case of the analysed data set will need further study, 
simply because in some wavelength regions it is consistently hard or
impossible to take a measurement.

In addition, for each measurement, an error value is also provided
from known instrumental characteristics and
uncertainty in calibration. 
 
\subsection{How many stellar populations?}
According to existing domain knowledge, it is likely
that there are (at least) two components of interest
\cite{kauff}. However, the 
first eigenvector explains more than 95\% of the data. 
Therefore, prior to deciding that a 2D representation space is
justified (i.e.  that at the given noise levels there is enough useful
information in the data and we are not attempting to model the noise
in a second component), we perform some simple, data-driven rank
tests.  We use the error matrix to derive thresholds for these tests.
\begin{figure}[ht!]
\begin{center}
\includegraphics[height=6cm]{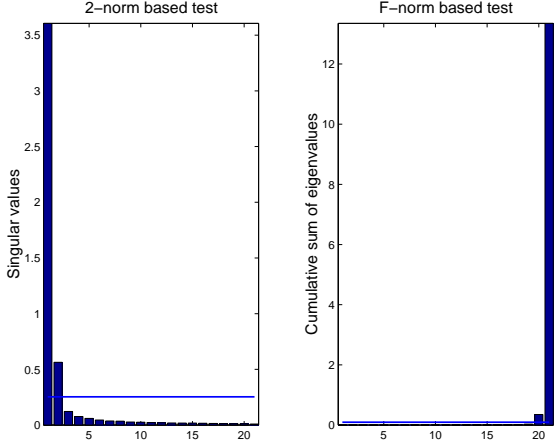}
\caption{Rank-tests for the matrix of stellar population spectra. The threshold values 
represent the 2-norm and the F-norm respectively of the matrix of known measurement errors.}
\label{ranktest}
\end{center}
\end{figure}
As shown on Figure \ref{ranktest},
both a 2-norm test and an F-norm test
\cite{Stewart} suggest that at the given error levels, the `clean' matrix 
of spectra has rank 2.  
Although it is known that these perturbation bounds
often tend to underestimate the rank \cite{Stewart}, there are no records of over-estimation, so we 
proceed to searching for a suitable two-dimensional latent representation space.

\subsection{The model}
Each stellar population spectrum is essentially a vector over a binned wavelength range, that represents the flux (in arbitrary units) per unit wavelength in the bin centered on the wavelength.  
The data matrix, having $T$ spectra in rows will be denoted by $\bd{Y} \in {\mathcal R}^{T\times N}$.
Single elements of this matrix will be referred to as $y_{tn}$, the $t$-th row is denoted by $\bd{y}_t$ or more explicitly $\bd{y}_{t,1:N}$ and likewise, the $n$-th column is denoted by $\bd{y}_n$ or $\bd{y}_{1:T,n}$. The whole matrix $\bd{Y}$ will also be referred to as $\bd{y}_{1:N}$.
Similar notational convention will also apply to other variables in the model. 

To account for the generation mechanism described in the previous section, namely that the observed optical spectrum of a galaxy is a linear superposition of the stars in the galaxy, a linear factor model will be assumed in this study. 
That is, we hypothesise that the $T$ observations can be explained as a superposition of $K<T$ latent underlying component spectra $\bd{s}_k \in {\mathcal R}^{K}$ (sometimes termed also as factors or hidden causes \cite{PPCA, Miskin, FastICA}) that are not observable directly but only through an unknown linear mapping $\bd{A} \in {\mathcal R}^{T \times N}$. Formally, this can be written as the following. 
\begin{equation}
\bd{y}_n = \bd{A}\bd{s}_n+\bd{\epsilon}
\label{model}
\end{equation}
In eq. (\ref{model}), the first term of the r.h.s. is the so called systematic component and $\bd{\epsilon}$ is the noise term or the stochastic component of this model. The noise term $\bd{\epsilon}$ is assumed to be zero-mean i.i.d. Gaussian, where the diagonal structure of the covariance accounts for the notion that all dependencies that exist in $\bd{Y}$ should be explained by the underlying hidden components. 

The $K$ components will be assumed statistically independent, this being a standard assumption of independent factor models \cite{FastICA, Miskin}. In the present application, this assumption is also cosmologically plausible, as there is little (no) interaction between stellar populations at different ages in a galaxy. The linearity of the mixture is physically justified as the fluxes of the hypothesised different subpopulations mix in an additive way.

\section{Independent projections of stellar population spectra}
A projection based approach is presented in this section.  This will
be accomplished in stages. A linear dimensionality reduction will
first be performed.  We then proceed at identifying independent
directions in the low-dimensional projection space. This multi-stage projection approach is
well-suited as a first attempt. It allows us to formulate sub-tasks in
statistical terms, and given the 2D nature of the problem, it also
allows us to benefit from a visual control over the data
representation obtained at various stages. 

\subsection{Dimensionality reduction using SVD}
Dimensionality reduction is a useful preprocessing stage for both
computational convenience and de-noising. It is well known from linear
algebra \cite{Golub, Stewart} that the best rank-K approximation of a
matrix under any unitarily invariant norm is its rank-K SVD (Singular
Value Decomposition) approximation. This is given by
$
\bd{Y}\approx\bd{UDV}^T
$, 
where $\bd{U}$ is the $T \times K$ matrix of left singular vectors,
$\bd{D}$ is the $K \times K$ diagonal matrix of singular values,
$\bd{V}$ is the $N \times K$ matrix of right singular vectors,
and $\bd{U}^T\bd{U}=\bd{V}^T\bd{V}=\bd{I}_K$, where $\bd{I}_K$ is the
$K$-dimensional identity matrix.
The projection is then  simply obtained as 
$\bd{X}=\bd{U}^T \bd{Y}$.

It should be pointed out that the scope of an SVD-based projection is
to identify an optimal (in the sense of any unitarily invariant norm)
subspace of the data space.  However, generally, individual singular
vectors or eigenvectors are not interpretable separately, as they
are not independent from each other. The same is true for PCA
\cite{PPCA}, for much the same reasons. 

\subsection{Finding non-orthogonal informative directions using contextual ICA}
We now turn to the key part of our analysis, where the directions of
independent projection need to be found. Approaches with this aim
are known under the name of Independent Component Analysis (ICA)
\cite{Aapo_One, FastICA, Nadal}. A vast number of ICA algorithms have
been developed over the last decade, each having different built-in
assumptions. In general terms, we can write the data likelihood of the
desired basis transformation as the following
\begin{equation}
\label{like}
p(\bd{x}_{1:N}|\bd{B})=\int d\bd{s}_{1:N} p(\bd{x}_{1:N}|\bd{s}_{1:N},\bd{B})p(\bd{s}_{1:N}).
\end{equation}
where $\bd{B}$ is the $K\times K$ unknown linear mapping (squared mixing matrix) that transforms the latent 
components $\bd{S}$ into $\bd{X}$. That is, as standard in ICA, 
instead of inferring $\bd{S}$ from $\bd{Y}$, it is easier to infer them from $\bd{X}$.

Assuming that the SVD projection performed as described in the previous 
section has removed the noise, then the noise term is a delta function 
\begin{equation}
p(\bd{x}_{1:N}|\bd{s}_{1:N},\bd{B})=\delta(\bd{x}_{1:N}-\bd{Bs}_{1:N})
\end{equation}
where $\bd{B}$ is a squared $K\times K$ parameter 
matrix (mixing matrix) that contains the 
desired new bases in its columns and $\bd{s}_{1:N}$ are the independent representations 
in the new basis --- both having to be estimated from the data. Thus,
(\ref{like}) reduces to the simple form below
\begin{eqnarray}
p(\bd{x}_{1:N})&=& |\det \bd{B}|^{-N} \prod_{k=1}^K p(s_{k,1:N})\\
&=&|\det \bd{B}^{-1}|^{N} \prod_{k=1}^K p((\bd{B}^{-1})_k\bd{x}_{1:N}).
\end{eqnarray}

Standard in squared ICA problems, it is easier to optimise for
the inverse of $\bd{B}$. That is, instead of the `top-down', or `generative' transform $\bd{B}$, 
we estimate the `bottom-up' or `projection' transform $\bd{B}^{-1}$.
However, without knowing $p(s_{1:N})$, this
is still an ill-posed problem.  Clearly, a mechanical application of any
ICA algorithm, out of the hundreds of existing ones, would produce
different results, although any of these would be somewhat arbitrary.
What we need is a well motivated prior distribution $p(s_{1:N})$.
However, as in most data mining applications of ICA, there is no such
information explicitly available.

\subsubsection{Exploiting correlations within the spectral data}

\begin{figure}[htb!]
\begin{center}
\includegraphics[width=0.9\hsize]{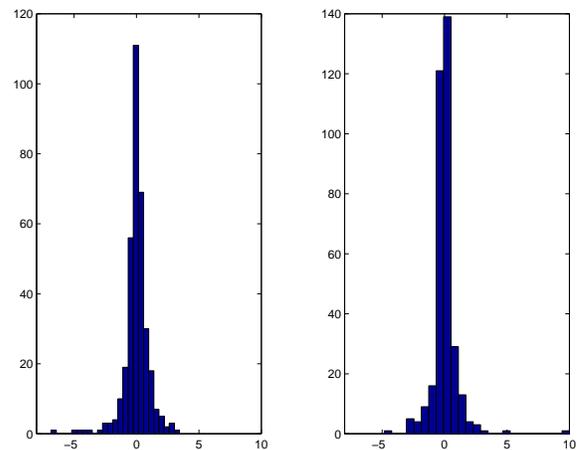} 
\caption{Histograms of the difference process. Kurtosis values are 11.0503 and 
33.9364 respectively.}
\label{xx_hist}
\end{center}
\end{figure}

Let us observe, however, that in spectral data, there is a natural
correlation structure between flux values in neighbouring wavelength
bins. This is what we exploit here, by capturing it in a form of a
contextual (predictive) model.
\begin{eqnarray*}
p(s_{k,1:N})&=& \prod_{n=1}^N  p(s_{kn}|s_{k,1:n-1})\\
&=&\prod_{n=1}^N  p(s_{kn}-E[s_{kn}|s_{k,1:n-1}])\\
&=&\prod_{n=1}^N p((\bd{B}^{-1})_k (\bd{x}_n-E[\bd{x}_n|\bd{x}_{1:n-1}]))
\end{eqnarray*}
$\forall k=1:K$. 
The advantage of doing so is that now we only need to specify the form
of density of the residual projections. Assuming a good enough
predictor, then the residual is likely to have a heavy tailed (termed
also super-Gaussian \cite{Aapo_One} or kurtotic) form of
density. Indeed, using just the simplest first order predictor, which
is an identity function
\begin{equation}
E[s_{kn}|s_{k,1:n-1}]\equiv s_{k,n-1}, \forall k=1:K
\end{equation}
the difference process $\bd{x}_n-\bd{x}_{n-1}$ of the data already
becomes highly kurtotic, as shown on Figure \ref{xx_hist}. 

A similar approach has been previously taken and successfully
demonstrated in the context of face image separation
\cite{Aapo_stoch}, where neighbouring pixel values of an image do also
exhibit significant correlations.

Let us denote $\bd{r}_{n}=\bd{x}_n-E[\bd{x}_{n}|\bd{x}_{1:n-1}]$ and 
$u_{kn}=s_{kn}-E[s_{kn}|s_{k,1:n-1}]$.
The data likelihood is then the following.
\begin{equation}
p(\bd{x}_{1:N})=|\det \bd{B}^{-1}|^N \prod_n \prod_k p((\bd{B}^{-1})_k\bd{r}_{n})
\end{equation}
where now we know that $p(u_{kn})$ is a super-Gaussian
density. Maximisation of this likelihood can now be accomplished by
employing any standard ICA algorithm --- over $\bd{r}_{1:N}$ rather
than $\bd{x}_{1:N}$. 
As the predictor
may not be very accurate (we just used an identity predictor in our
experiments),  it is preferable to chose a robust
approximation of the generalised exponential density, that grows relatively
slowly in $|u_{kn}|$. Following the arguments in \cite{Aapo_One}, in
our experiments we have used the following:
\begin{equation}
\log p(u_{k}) \propto \exp(-u_k^2/2),
\end{equation}
and the optimisation has been performed using the Newton method
implemented in the FastICA routines \cite{FastICA},  employing the
faster deflationary approach. This has a cubic 
convergence \cite{Aapo_One}. Indeed, highly kurtotic independent
projections have been found (kurtosis: 33.8796 and 12.7484
respectively) on the data investigated, in about ten iterations only.

\begin{figure*}[htb!]
\begin{center}
\includegraphics[width=17cm, height=5cm]{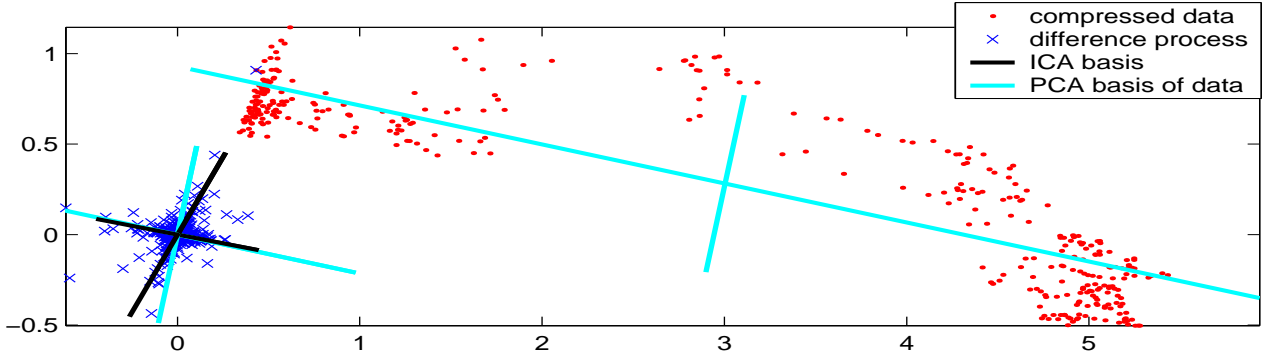}
\caption{Geometric illustration of the described ICA procedure. The number of 
points shown as dots equals the number of different wavelength bins in
the set of measurements, each point being the 2D compressed
representation of the fluxes at one of these bins. The differences
between these 2D vectors at consecutive wavelength bins are marked
with '$\times$'. The PCA basis of the data is
superimposed (in light color) and also translated to the origin for comparison with the
ICA basis found on the difference process. The ICA basis is shown in dark color.}
\label{geom}
\end{center}
\end{figure*}

A geometric illustration of the procedure just described 
is shown on Fig.~\ref{geom}. 
The SVD-compressed data are shown as dots
and indeed, informative directions would be difficult to determine
directly from the data. The scatter-plot of the difference process is
shown as crosses. A star-like structure is apparent,
with two main, non-orthogonal linear directions of high data
density. These are the new bases (columns of $\bd{B}$) that are 
determined by the ICA procedure. 
Indeed, the two directions defined by the new bases found by the
algorithm are highlighted on the plot as dark lines. The PCA axes of
the data are also shown on the same plot for
comparison. Interestingly, one of the axes is almost identical to one
of the independent directions. The second is, however, just orthogonal
to the first, while the ICA axes are not orthogonal to each other but
do follow the two main directions of high density in the data.

To obtain the component spectra from the ICA procedure described, we
simply compute the projections of the individual flux values of all
galaxies at all wavelength bins onto the new bases (which is the
composition of the two linear transforms performed during the analysis
process described):
\begin{equation}
\bd{s}_{1:N}=\bd{Bx}_{1:N} = \bd{Ay}_{1:N}
\end{equation}
where $\bd{B}$ and $\bd{A}$ now denote the recovered mixing parameters. Specifically, after $\bd{B}$ 
is found, $\bd{A}$ is computed as the matrix product $\bd{U}\bd{B}$.

\begin{figure*}[ph!] \label{reco}
\begin{center}
\includegraphics[height=6cm]{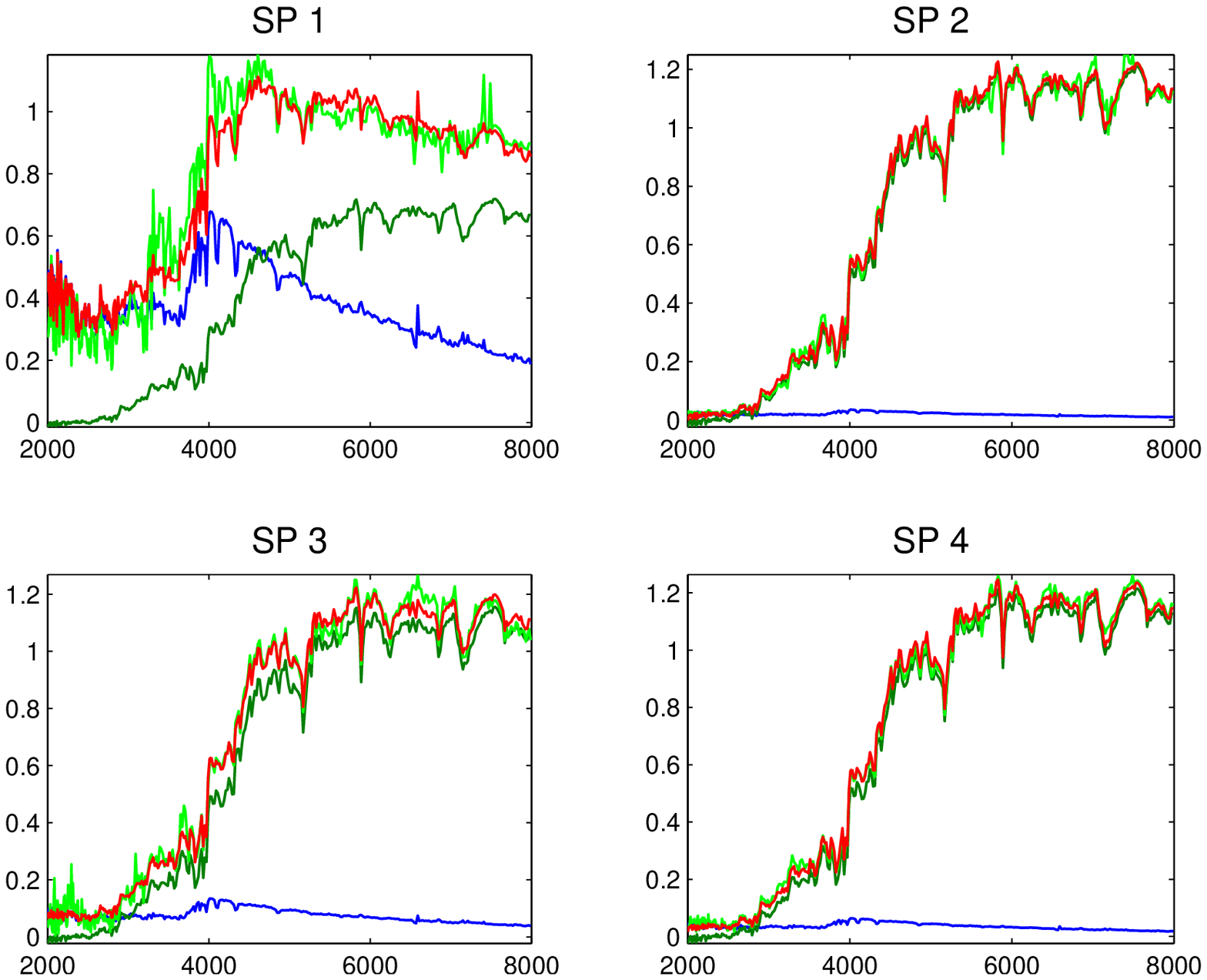}
\includegraphics[height=6cm]{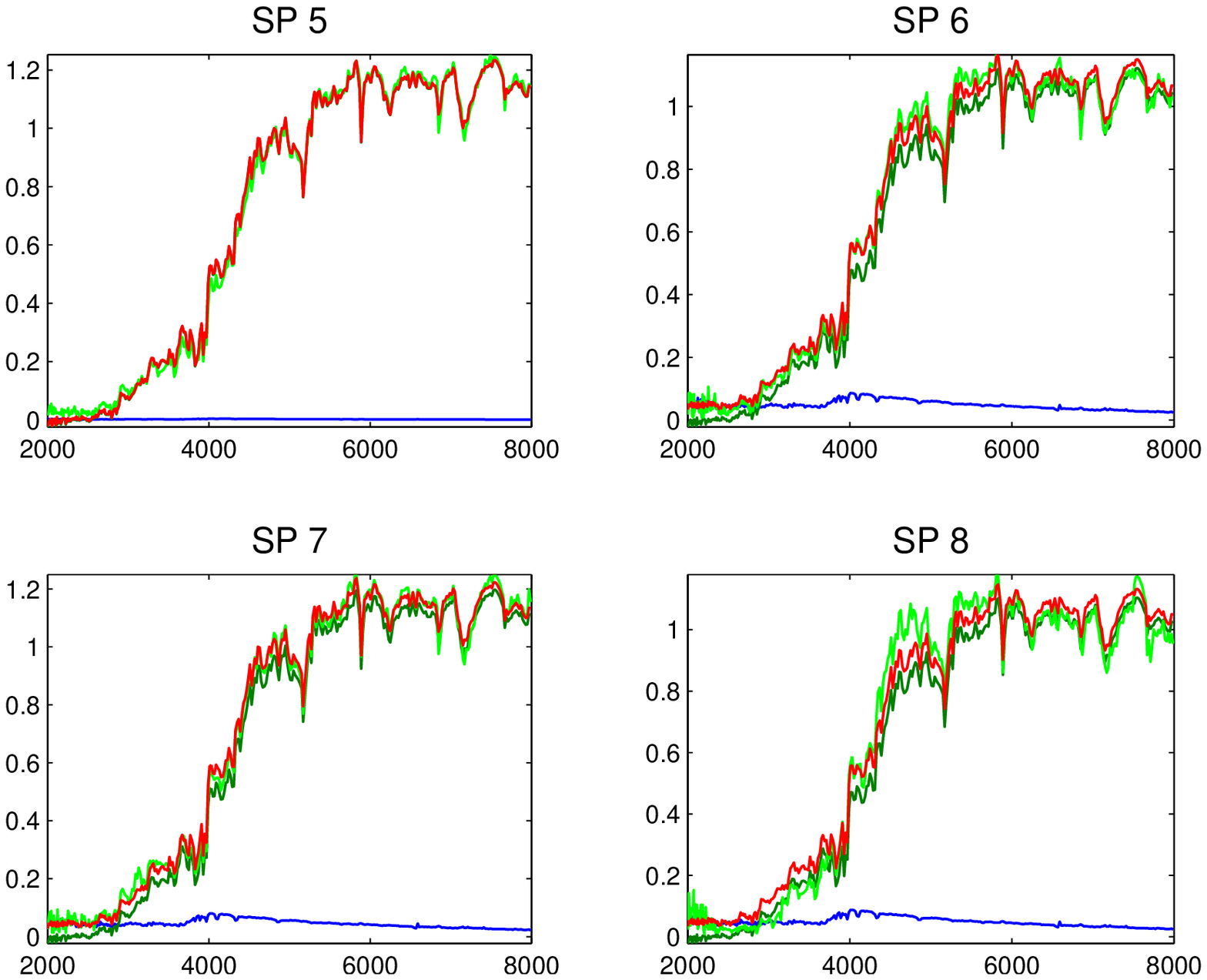}
\includegraphics[height=6cm]{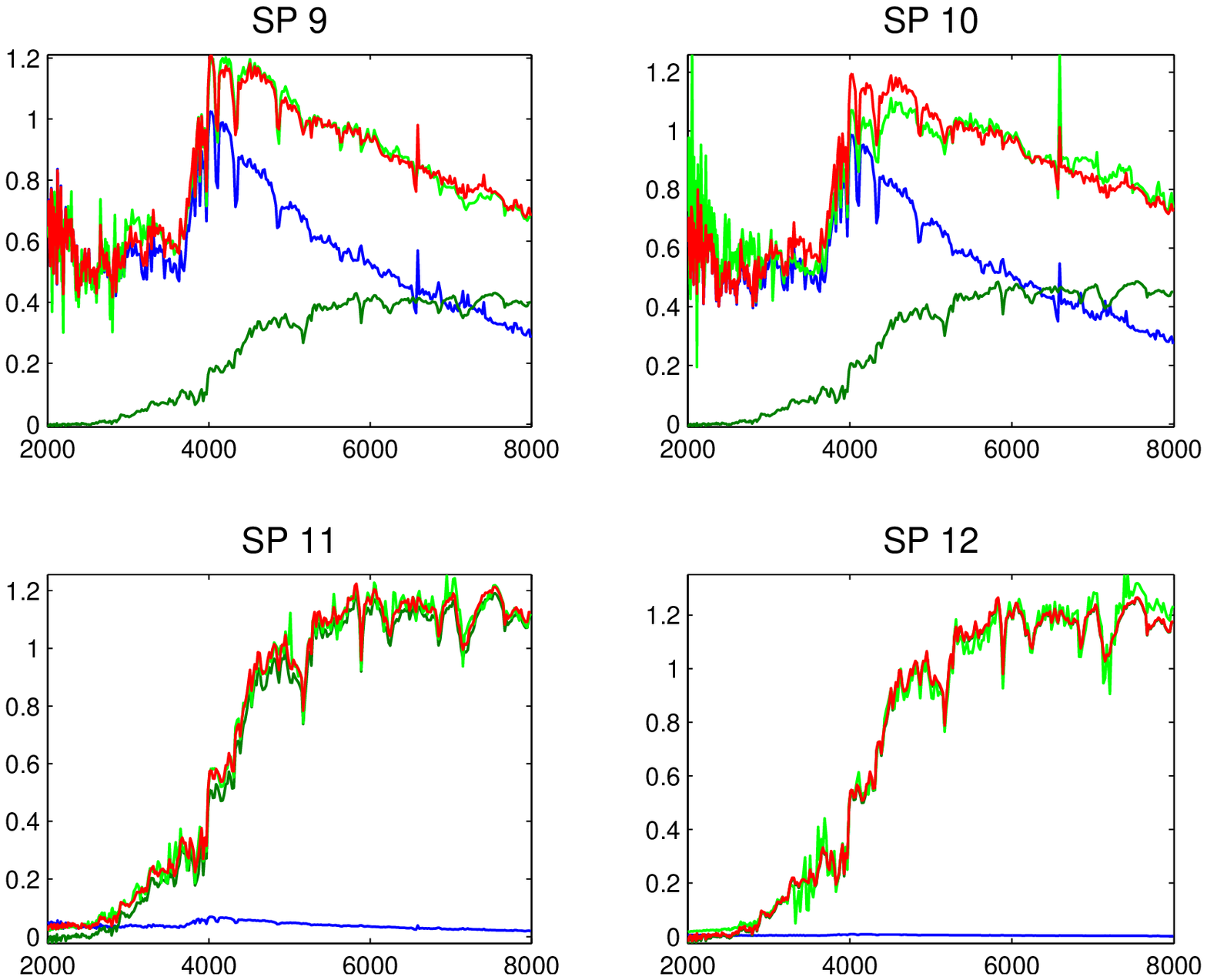}
\includegraphics[height=6cm]{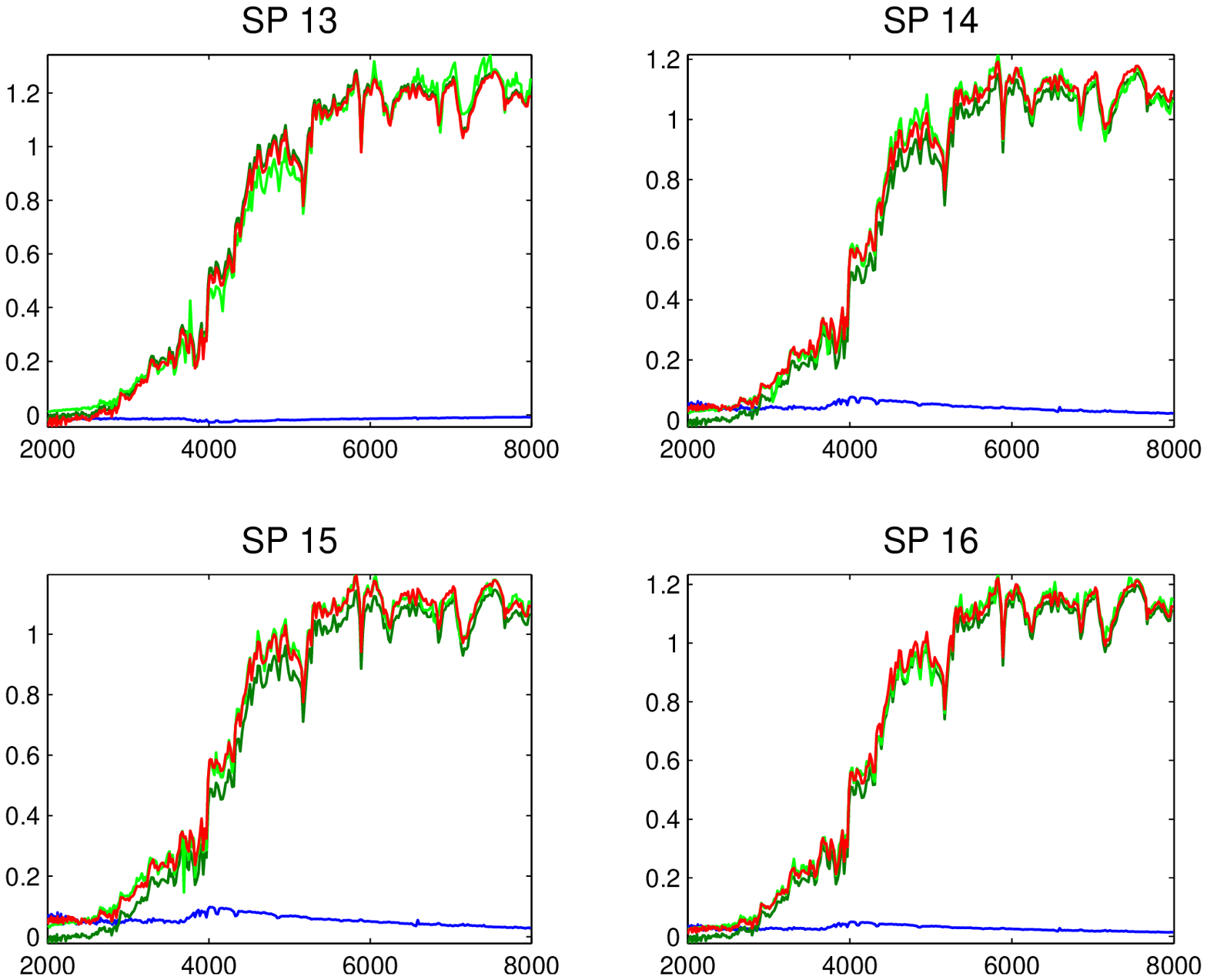}
\includegraphics[height=6cm]{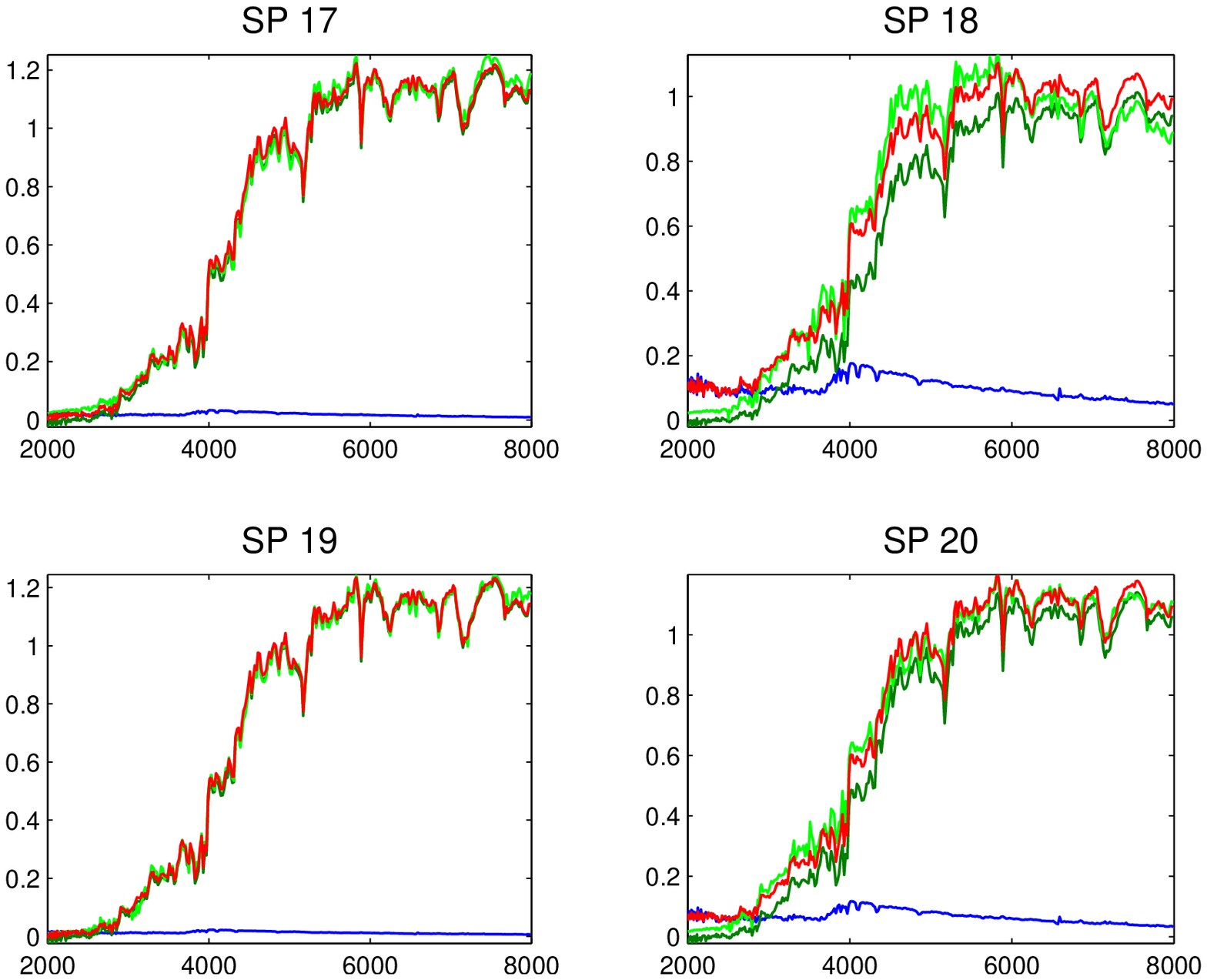}
\includegraphics[height=2.7cm]{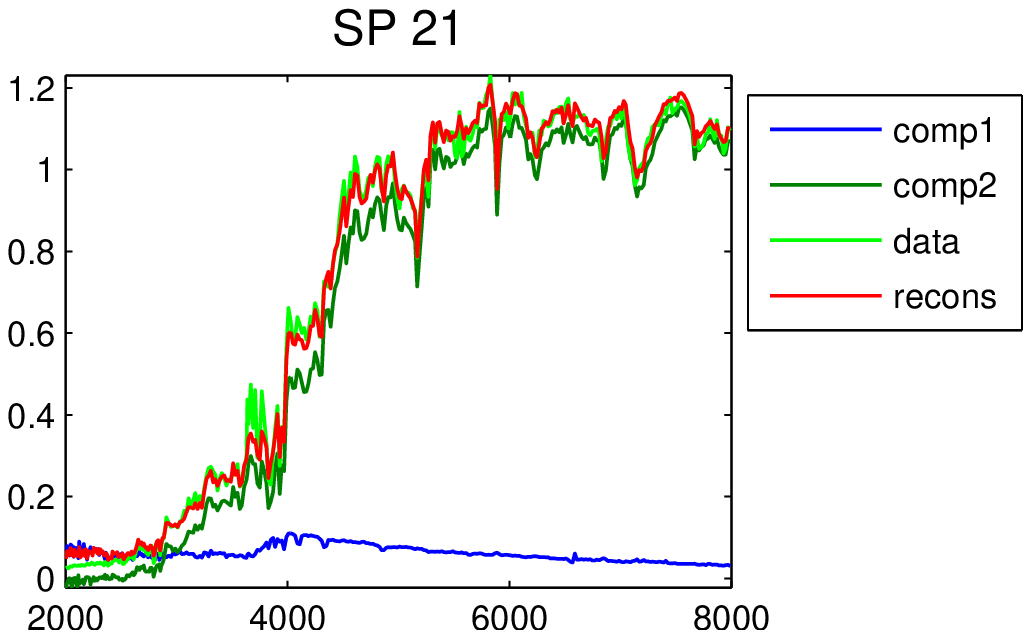}
\caption{The reconstruction of stellar population spectra 
using the projection based contextual ICA. Here, our algorithm has
decomposed the observed spectrum of each galaxy into two inferred
populations (blue and dark green --- the two darkest colors, in black-and-white printing), whose sum is given in red, superimposed with the data in light green (intermediate grey and light grey respectively, in black-and-white printing). 
In many cases (e.g. \#1,9,10,18), a significant younger stellar
population is found, which is confirmed by our detailed physical
modelling. The numbering of the spectra corresponds to the enumeration order of the corresponding galaxies as given in Section 2.}
\label{recons}
\end{center}
\end{figure*}

The component-wise 
reconstruction of the 21 individual stellar 
population spectra 
from their 
independent components are shown on Figure \ref{recons}. 
The physical 
interpretability of these components will be assessed in Section 5, however, as a data-driven 
observation, it is interesting to note 
that the recovered spectral components turned out to be positive valued ---
although positivity has not been artificially imposed at
this stage during the analysis process. We note that indeed negative values of 
the flux would be difficult to interpret, therefore in the next section we discuss a different approach,
where the required latent density $p(\bd{s}_{1:N})$ is derived from a positivity constraint.

%%%%%%%%%%%%%%%%%%%%%%%%%%%%%%%%
\section{A Positivity-based approach}

The use of positive factorisation of positive matrices to replace PCA
for analysing positive data, such as spectral data dates back to work
reported in \cite{juvela}.  Positive (more exactly non-negative)
factor models have been further developed in
\cite{Lee_Seung}. However, in the absence of either a density-based or
a geometric interpretation, the implicit assumptions are not clear and
therefore the interpretation of the results may not be straightforward.
Somewhat related, in \cite{Miskin}, a fully Bayesian
formulation of a positive factorisation model is given and variants
with sparse positive priors are applied to synthetic stellar
population spectra in a different context of investigations than ours. 
Retaining the probabilistic framework, that allows us 
to make all assumptions explicit, we present a simpler version
of their algorithm, based on maximum a posteriori (MAP) / maximum
likelihood (ML) estimation, which will highlight the link with
positive factorisation algorithms \cite{Lee_Seung}. A MAP or ML
estimation is sufficient for our purposes as we are concerned with a
data explanation task for a fixed data set only. Once the most
appropriate model is found, the full Bayesian machinery remains
available to derive fully generative models that are able to better
generalise on new data.
  
Retaining the positivity of the representation, and the fact that the
overall transform in the approach adopted in the last section has been
linear, the following linear model can be formulated.
\begin{equation}
p(\bd{y}_n|\bd{A},\bd{E}_n)=\int d\bd{s}p(\bd{y}_n|\bd{As},\bd{E}_n)
p(\bd{s})
\end{equation}
A Gaussian measurement noise will be assumed. Furthermore, according
to prior knowledge about the levels of imprecision of the physical
instruments, which vary independently for each stellar population and
each wavelength bin, we will have individual diagonal variances
$\bd{E}_n^2$ at each wavelength bin $n$, 
$p(\bd{y}_n|\bd{s},\bd{A},\bd{E}_n)\sim {\mathcal N}(\bd{y}_n|\bd{As}_n, \bd{E}_n^2)$. 
In rest, we use the same 
notation as before, $\bd{y}_n$ refers to the 
relative flux values observed at the $n$-th wavelength bin, $\bd{A}$ is the unknown 
mixing matrix parameter and $p(\bd{s})$ is the distribution of the latent components.

Now the latent prior needs to be specified. Apart from its positive
support we don't have much information in this respect. Therefore we
formulate a vague exponential prior, that is
\begin{equation}
p(\bd{s}) \propto \prod_k \exp(-\alpha_{kn} |s_{k}|)
\end{equation}
where $\alpha_{kn}=\alpha, \forall k,n$ is a small positive
constant. If $\alpha \rightarrow 0$, then the prior becomes
non-informative but also improper and in this case the MAP estimation
procedure given below becomes ML.

To obtain a MAP estimate, the posterior needs to be maximised 
$p(\bd{s}_n|\bd{y}_n,\bd{A},\bd{E}_n)
\propto p(\bd{y}_n|\bd{s}_n,\bd{A},\bd{E}_n)
p(\bd{s}_n)$ which is proportional to the complete data likelihood. 

Positivity of the elements of both $\bd{A}$ and $\bd{S}$ are then imposed by 
adding Lagrangian terms \cite{opt} to the complete data log likelihood. 
\begin{eqnarray*}
{\mathcal L}&=& \sum_n \ny \log p(\bd{y}_n|\bd{s}_n,\bd{A},\bd{E}_n)+\log p(\bd{s}_n) \zr\\
&+&Tr \bd{L}_1^T \bd{A} +Tr \bd{L}_2^T \bd{S}
\end{eqnarray*}
where $\bd{L}_1$ and $\bd{L}_2$ are a set of non-negative Lagrange multipliers
and $Tr$ denotes the trace of a matrix.

From the stationary equations w.r.t. $\bd{A}$ and $\bd{S}$ the Lagrange multipliers
are obtained.
\begin{eqnarray*}
\bd{L}_1&=&\sum_n \bd{E}_n^{-2}\bd{As}_n\bd{s}_n^T - \sum_n \bd{E}_n^{-2}\bd{y}_n\bd{s}_n^T\\
\bd{L}_2&=&\sum_n \bd{A}^T\bd{E}_n^{-2}\bd{As}_n - 
\sum_n \bd{A}^T\bd{E}_n^{-2}\bd{y}_n
+\alpha
\end{eqnarray*}
Now from the 
Karush-Kuhn-Tucker conditions \cite{opt} $L_{tk}A_{tk}=0$ and $L_{kn}s_{kn}=0, \forall t,k,n$,
we have two fixed point equations which provide the convergent alternating iterative algorithm of the multiplicative form below.

\begin{eqnarray*}
\bd{A}&=& \bd{A}\odot (\bd{E}^{-2}\odot \bd{Y})\bd{S}^T  \oslash
 [\bd{E}^{-2}\odot \bd{AS}]\bd{S}^T \\
\bd{S}&=& \bd{S}\odot\bd{A}^T(\bd{E}^{-2}\odot \bd{Y})  \oslash
\ny \bd{A}^T[\bd{E}^{-2}\odot \bd{AS}] +\alpha\zr
\end{eqnarray*}
where $\odot$ denotes element-wise multiplication and $\oslash$ denotes element-wise division.
If $\alpha=0$, the iterative algorithm above is identical to the
least-squares based non-negative factorisation algorithm proposed in
\cite{Lee_Seung} from a non-probabilistic starting point, with the
only difference that now we also have the $\bd{E}$ terms to account for known
measurement errors.

\section{Evaluation}
\subsection{Data driven evaluation} 
Here we assess the effectiveness of
the methods presented above according to two indicators: 
(1) data reconstruction and (2) the 
mutual information (MI) between the components 
of the representation created. 

\begin{table}[htb!]
\begin{center}
\begin{tabular}{|c|c|c|c|} \hline
Method & Min & Median & Max\\
\hline
SVD, cICA & 0&$3.36 \times 10^{-4}$&0.294\\
NMF &  $10^{-10}$&$3.64 \times 10^{-4}$& 0.317\\
NMFe & 0& $4.08 \times 10^{-4}$& 0.351\\
\hline
\end{tabular}
\caption{Data reconstruction errors under the L2 norm. cICA = contextual ICA, NMF = Non-negative matrix factorisation with $\alpha$=0,
NMFe = NMF with exponential prior having $\alpha = 0.1$}
\end{center}
\end{table}

Table~1 shows the data reconstruction results across all $N\times T$
measurements for the various methods. These are measured as the squared distances between the
data and the reconstruction, which is in accordance with the Gaussian
noise assumption.  The reconstruction error of cICA are identical to
that of the SVD, since the ICA transform does not reduce the dimensionality further.
For NMF and NMFe, 15 randomly initialised runs were performed and the one with the highest likelihood value was selected
in this evaluation, in order to avoid the possibility of getting trapped in a local optimum. Indeed, note that the positivity-based single stage approach involves a non-convex optimisation whereas the SVD-based preprocessing is a 
convex problem. 
The median of the error appears to be the smallest in the case of SVD+cICA, however,
a pairwise application of the non-parametric Wilcoxon rank sum test to
the whole sample distribution of the individual $T\times N$ reconstruction errors returned that the difference of the 
medians for cICA and NMF is not statistically significant at the 0.05\%
level. In the case of all other pairs, the differences between medians
were found statistically significant --- as expected, the additional term that enforces the latent prior 
distribution in NMFe causes a slight increase in data reconstruction error. However, this small difference on its own would not be a practically concerning issue here, as we deliberately
formulated constraints in our models in order to obtain
representations that are statistically independent to a higher degree
--- in the hope that these may then be interpretable and 
capable of being independently further
processed. More interesting is therefore to evaluate to what extent do these methods achieve statistically independent 
representations.  

The information theoretic quantity that measures the degree of statistical
dependence is the Mutual Information (MI) \cite{Cover}.  This is a
non-negative value that vanishes if the component densities are
perfectly independent (the smaller the MI the better). It can be shown
\cite{Nadal} that maximising the ICA log likelihood in the noise-free case 
is equivalent to minimising MI between the representation components.  Here we compute 
the sample-based MI of the components,
as estimated according to the procedure described in \cite{MI}. 
The comparative values obtained by various methods for this data are shown in Table~2.
For the contextual ICA method, that works on
predictive residuals, two values are given: the first value is
computed from the projections of the residuals whereas the second
value is computed from the projections of the data, taken as it was
iid. (just for obtaining a value to compare with those obtained with the rest of the methods).
\begin{table}[htb!]
\begin{center}
\begin{tabular}{|c|c|} \hline
Method & MI\\
\hline
NMF & 0.5923\\
NMFe & 0.6019\\
cICA & (u) \bd{0.00010394}\\
&  (s) \bd{0.5583}\\
PCA & 1.0931\\
\hline
\end{tabular}
\caption{Sample-based mutual information estimates of the two components obtained 
with the various methods for the set of stellar population spectra
investigated. Smaller values signify higher independence achieved after the estimated basis transform.}
\end{center}
\end{table} 
It is apparent from the table that in both cases the contextual ICA
model achieves lower MI (greater independence of the components).  The
positivity-based method with non-informative improper prior is the next
best performing method, whereas employing sparse priors (greater
$\alpha$ values) does not lead to components that are more independent,
for this data. PCA is used as a baseline, as we know that it doesn't
produce an independent representation. 
Visually, the positivity based components do not look much
different from the ones obtained from cICA (Fig.~\ref{recons}), although 
they are slightly more noisy. The PCA-based
decomposition, however, in general, are theoretically not guaranteed to be interpretable, as already 
discussed in Section 3.1 (see \cite{PPCA} for details).
We now turn to evaluate the interpretability of the obtained results from the 
astrophysical perspective.

\subsection{Astrophysical evaluation}

Here, we compare the results from the linear independent basis
transformation analysis with an entirely independent determination of
the star formation history, based on detailed astrophysical models
of the evolution of stellar populations.
 
The 21 observed spectra have been analysed by matching them with
synthetic stellar population spectra.  For each of the observed
spectra, a two-stellar population component model \cite{jimenez} was
fitted. The age, chemical abundance and relative mass fraction of each
component were allowed to vary freely. The best fit in each case was
determined by a minimum $\chi^{2}$ test \cite{nol1,kraft}.

\begin{figure*}[ht!]
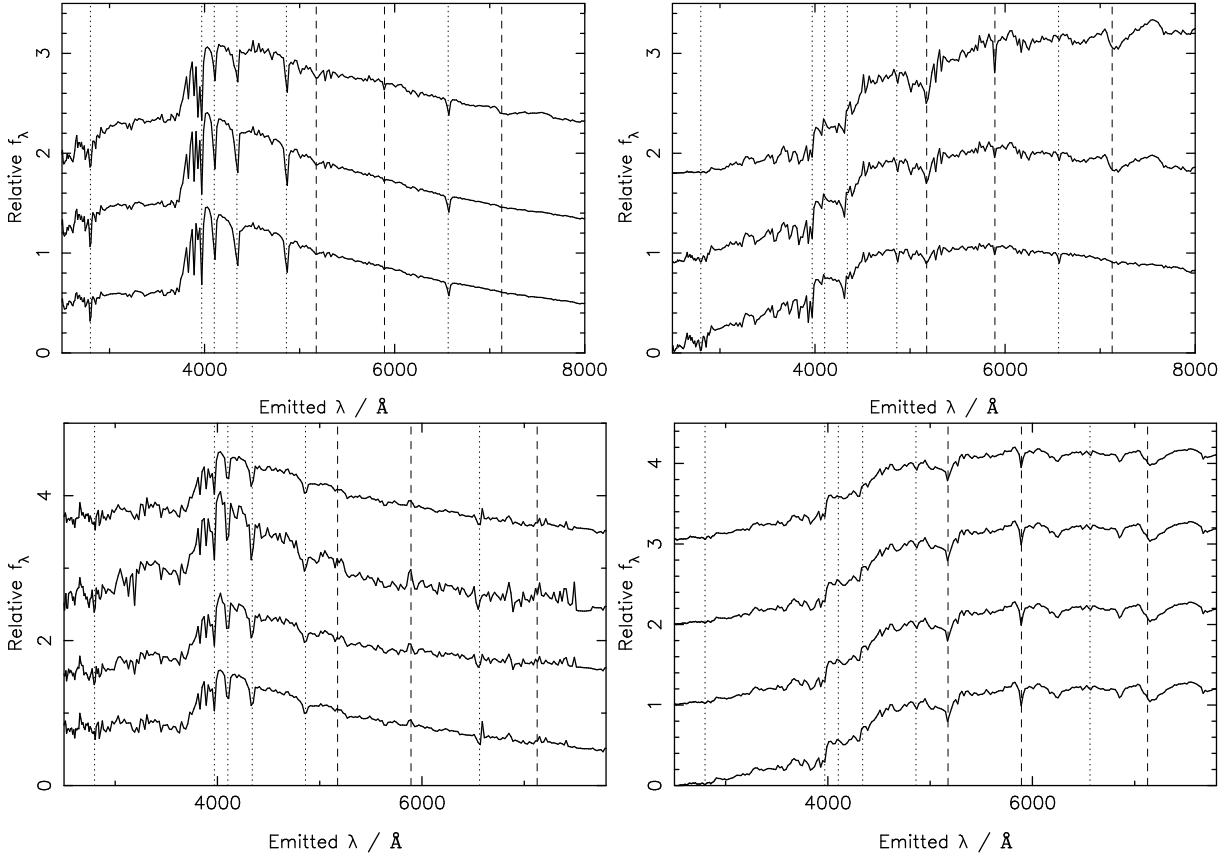

\begin{center}
\includegraphics[height=8cm,angle=-90]{jimenezspectra-0.7ica.ps}
\includegraphics[height=8cm,angle=-90]{jimenezspectra-10ica.ps}
\includegraphics[height=8cm,angle=-90]{comp1.ps}
\includegraphics[height=8cm,angle=-90]{comp2.ps}
\caption{Comparison of the derived components with physical models
of the stellar population spectra.
{\it Top two plots:} 
Synthetic stellar population spectra according
to the physical models of \cite{jimenez}. {\sl Right:} 
Spectra of a population of age 10~Gyr, 
where chemical abundance, from bottom to top, 0.2, 1.0 and 2.5 times
solar; {\sl Left:} age = 0.7~Gyr, same chemical abundances. The dotted
lines mark some of the absorption features in the spectrum which are
typically strong in young stellar populations, and the dashed lines
mark some of the absorption features which are typically strong in old
stellar populations. From left to right, the absorption line species are: MgII (2799 \ang), H$\varepsilon$ (3970 \ang), H$\delta$ (4102 \ang), H$\gamma$ (4340 \ang), H$\beta$ (4861 \ang), Mgb (5175 \ang), NaD (5893 \ang), H$\alpha$ (6563 \ang), TiO (7126 \ang). {\it Bottom two plots:} the 2 components found from the
various different linear independent basis transformation analyses,
from bottom to top: cICA, NMF, NMFe, PCA. (The spectra are shifted along the vertical axis for the sake of clarity.) The recovered spectra are
convincingly disentangled into one component with young stellar population features (MgII, H$\varepsilon$, H$\delta$, H$\gamma$, H$\beta$, H$\alpha$: dotted lines) and
shape, and a second with the features (Mgb, NaD, TiO: dashed lines) and shape of an old,
high chemical abundance stellar population.}
\label{modelspec}
\end{center}
\end{figure*}

The principle for creating synthetic stellar population spectra is
simple, although the input physics is complex. The spectral energy
distribution of a star evolves according to its initial mass and
chemical abundance. If the initial mass distribution and the chemical
abundance of a stellar population is known, and the spectral evolution
of each individual star in this initial population may be modelled,
the stellar spectra may be summed over the mass distribution at any
point in time to give the integrated spectrum of the population at
that age. The ingredients for a stellar population model are
therefore: stellar evolutionary tracks; a library of stellar spectra;
a method of calibrating the theoretical luminosity and effective
temperature, determined from the evolutionary sequence, so that the
appropriate atmosphere may be assigned to each star at each time-step
in its evolution. In contrast, the linear independent basis
transformation method employs no knowledge of the underlying physics
in the observed spectra.

Model spectra \cite{jimenez} for a young (0.7 Gyr) and an old (10 Gyr)
stellar population, with three different chemical abundances (0.2, 1.0
and 2.5 times solar abundance) are shown in Fig.~\ref{modelspec}, together
with the spectra recovered from the linear independent basis
transformation analyses. 
The similarity is most apparent. The data-driven linear analyses,
whilst not reproducing any of the physical model spectra precisely
(which would not be expected anyway), extract many of the important
identifying characteristics of these two categories of model spectra, which are
indeed quite different from each other both in their overall shape and 
details. Some of the most important features, 
the so-called absorption-lines, are marked with dashed (Mgb, NaD and TiO, typically strong features in old, high chemical abundance stellar population spectra) and dotted (MgII, H$\varepsilon$, H$\delta$, H$\gamma$, H$\beta$ and H$\alpha$, typically strong in the spectra of young (\ls\ 1~Gyr) stellar populations) 
vertical lines on the figure (Fig.~\ref{modelspec}), which demonstrate the physical 
interpretability of the representations created by the basis transformation analyses.

\begin{figure*}[htb!]
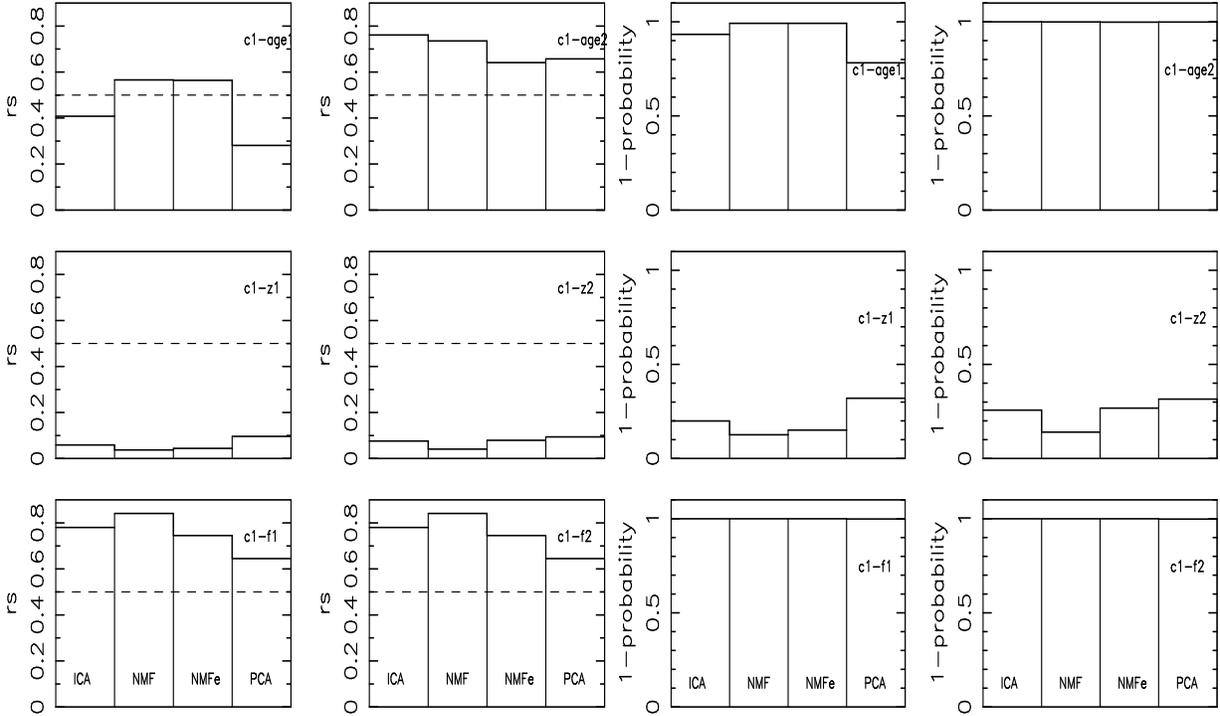

\begin{center}
\includegraphics[width=9.5cm,height=8cm,angle=-90]{histcorr.epsf}
\includegraphics[width=9.5cm,height=8cm,angle=-90]{histprob.epsf}
\caption{{\it Left:} 
The modulus of Spearman's rank order correlation coefficient $rs$ for
the weight of the first component of the various linear basis
transformation analyses, $c1$, correlated with the parameters recovered
from the physical analysis i.e. a two-component spectral model fitting to the observed galaxy
spectra. The parameters investigated in this correlation analysis are
the age, chemical abundance $z$ and relative mass fraction
$f$ for each component. Values greater than 0.5 indicate a strong
correlation. {\it Right:} the significance of $rs$. High values of
(1-probability) indicate a high level of significance.}
\label{spear}
\end{center}
\end{figure*}

\begin{figure*}[htb!]
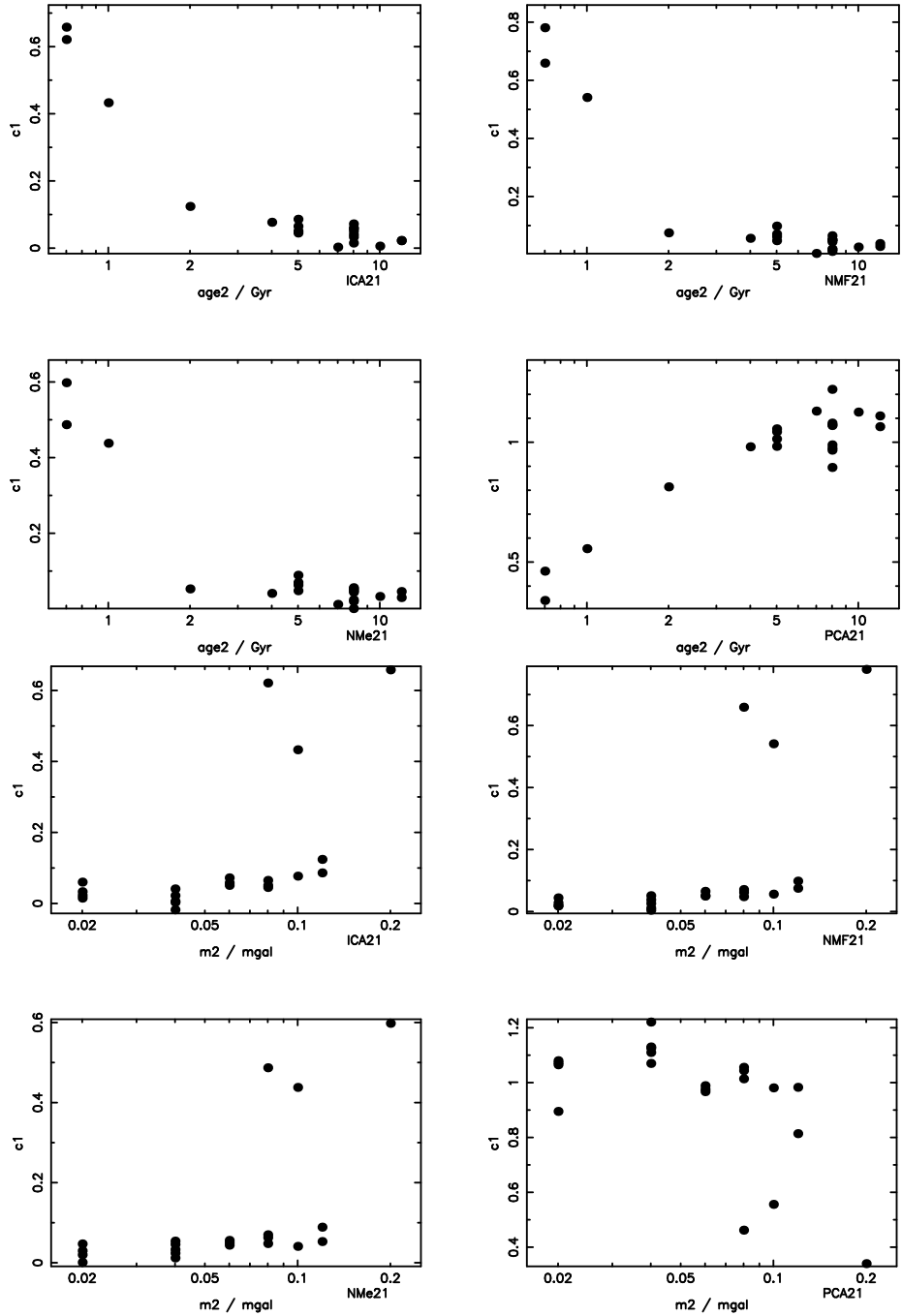

\begin{center}
\includegraphics[width=9cm,angle=-90]{c1-age2.epsf}
\includegraphics[width=9cm,angle=-90]{c1-f2-msort.epsf}

\caption{Scatter-plots showing the correlation of {\it (top)}
the age of the younger
stellar population and {\it (bottom)} the mass fraction of the smaller
stellar population determined from the model spectra fitting with the
weight of the first component of the various linear basis
transformation analyses (c1). A high value (low for PCA) of c1 clearly
corresponds to a substantial young (\ls\ 1 Gyr) stellar population.}
\label{corrplots}
\end{center}
\end{figure*}

We correlate the star formation history parameters derived from
fitting the two-component model spectra to the observed spectra 
(i.e. a physical analysis approach) with
the weight of the contributions from the linear basis transformation
analyses, by defining these weights as $c_k= a_{tk}/\sum_{k'}a_{tk'}$ for any 
given spectrum $t$. Here, $a_{tk}$ is the $(t,k)$-th element of the matrix $\bd{A}$ of 
the new basis and $k=1:2$. 
Fig.~\ref{spear} shows the results of the
correlations, and Fig.~\ref{corrplots} 
graphically shows some of these correlations.

From Figs.~\ref{spear} and \ref{corrplots} we can conclude that, for the ICA and NMF
analyses, 
c1 (and hence also 
c2, as $c_1+c_2=1$)
correlates with the proportion of young (\ls\ 1~Gyr) 
stellar population component present in the observed spectrum, regardless of their chemical abundance.

\section{Conclusions}

We have presented a scientific data mining application that searches
for linear independent basis transformations of galaxy spectra to find
the spectra of individual stellar populations characterised by age and
chemical abundance. We have shown that characteristic stellar
population components of elliptical galaxies can be disentangled from
the observed 
spectra of these galaxies, without the use of
detailed physical models. The components returned by the linear basis
transformation analyses are clearly physically interpretable, with one
component displaying the shape and many of the absorption-line
features typical of a young stellar population, and the second
component having the over-all shape and typical absorption features of
an old, high chemical abundance stellar population. The weights of the
contributions from the linear basis transformation analyses correlate
well with both the ages of the younger stellar populations and the
mass fractions of the smaller stellar populations determined from the
(completely independent) detailed physical modelling of the observed
galaxy spectra.

The computational demand
of the projection approach presented is essentially that of the SVD
computation, so it is expected that the method is easily applicable to large
sets of measurements as they become available. The positivity based
approach, per iteration, has a comparable scaling, however, in all our
experiments the number of iterations to convergence was of an order of
magnitude larger for the positivity based single stage
approach. Further study is necessary to investigate models with other
types of positively supported priors as well as refining the best
performing models and algorithms to be able to deal with previously
unseen data.

The use of the data
analysis presented in this paper, integrated with the more complex
process of astrophysical analysis will be detailed elsewhere
\cite{nolan-future}.  
We intend to investigate the effectiveness of these data-driven
methods on larger sets of UV-optical spectra as they become available,
where more comprehensive statistical evaluation will be possible. 
From  the analysis of large archives of
galaxy spectra using this technique, we hope to address some of the
fundamental questions in astrophysics, that of when and how galaxies
form and evolve.

\section*{Acknowledgement}
This research was partly supported by a Paul \& Yuanbi Ramsay research award 
from the School of Computer Science of the University of Birmingham.


\begin{thebibliography}{99}
\bibitem{Cover} T.Cover and J. Thomas, 
Elements of Information Theory. John Wiley and Sons, Inc.,1991.
\bibitem{MI}
G.A. Darbellay and I. Vajda,
Estimation of the information by an adaptive partitioning of the observation space. 
IEEE Trans. Information Theory, Vol. 45, no. 4, pp. 1315-1321, May 1999.  
\bibitem{eggen} 
O.J. Eggen, D. Lynden-Bell, A.R. Sandage, Evidence from the motions of old stars that the Galaxy collapsed, Astrophys. J., 136, 748, 1962
\bibitem{Golub} G.H. Golub and C.F. Van Loan, Matrix Computations.
Johns Hopkins University Press, 1989.
\bibitem{Aapo_stoch}
A. Hyvarinen, Independent Component Analysis for Time-dependent Stochastic Processes,
Proc. Int. Conf. on Artificial Neural Networks (ICANN'98) 1998, pp. 541--546.
\bibitem{FastICA} A. Hyvarinen, Fast and Robust Fixed Point Algorithms 
for Independent Component Analysis. IEEE Transactions on Neural Networks
10(3): 626--634, 1999.
\bibitem{Aapo_One} A. Hyvarinen, One-unit Contrast Functions for Independent Component Analysis: A Statistical Analysis. 
Proc IEEE Neural Networks for Signal Processing VII, 1997,
pp. 388--397. 
\bibitem{jimenez} R. Jimenez, P. Padoan, F. Matteucci, A. Heavens,
Galaxy formation and evolution: low-surface-brightness galaxies 
Mon. Not. R. Astron. Soc. 299, 123, 1998
\bibitem{juvela} M. Juvela, K. Lehtinen, P. Paatero, The Use of Positive Matrix Factorisation 
in the analysis of Molecular Line Spectra. Mon. Not. R.
Astron. Soc., 280 pp.616--626, 1996.
\bibitem{kauff} G. Kauffmann, S. White, B. Guideroni, 
The Formation and Evolution of Galaxies Within Merging Dark Matter Halos,
Mon. Not. R. Astron. Soc. 264, 201, 1993
\bibitem{kraft} R P Kraft, L A Nolan, T J Ponman, C Jones, S Raychaudhury, 
A Chandra observation of the nearby lenticular galaxy
NGC 5102: where are the x-ray binaries? Astrophys. J., submitted, 2004
\bibitem{larson} R. B. Larson, Models for the formation of elliptical galaxies,
 Mon. Not. R. Astron. Soc. 173, 671, 1975
\bibitem{Miskin} J. Miskin, Ensemble Learning for Independent 
Component Analysis. PhD Thesis. University of Cambridge, 2000.
\bibitem{Nadal} T.-W. Lee, M. Girolami, A.J. Bell and T.J. Sejnowski, 
A Unifying Framework for Independent Component Analysis.
Computers and Math. with Applications, vol. 39, no. 11, pp.1--21, 2000.
\bibitem{nol1} L.A. Nolan et al., The star-formation history of galaxies in the GEMS groups, in preparation.
\bibitem{nolan-future} L.A. Nolan, M.O. Harva, A. Kab\'an, S. Raychaudhury, Finding Young Stellar Population
in Elliptical Galaxies from Independent Components of their UV-optical spectra, in preparation for submission to
Mon. Not. R. Astron. Soc. 
\bibitem{PPCA} M. Tipping and C. Bishop, 
Probabilistic Principal Component Analysis,
Journal of the Royal Statistical Society, Series B,61,
 Part 3, pp. 611--622, 1999.
\bibitem{Lee_Seung} D. Lee and S. Seung, 
Algorithms for Non-Negative Matrix Factorisation,
Advances in Neural Information Processing Systems 13, 556--562, MIT Press, 2001.
\bibitem{Stewart} G. W. Stewart,
Perturbation Theory for the Singular Value Decomposition.
Appeared in SVD and Signal Processing, II, R. J. Vacarro ed., Elsevier, 1991.
\bibitem{opt} H.A. Taha, Operations Research -- An Introduction. 1997, Prentice-Hall.
\bibitem{missing} O. Troyanskaya, M. Cantor, G. Sherlock, P. Brown, T. Hastie,
R. Tibshirani, D. Botstein and R.B. Altman, Missing Value Estimation Methods for DNA
Micro-arrays. Bioinformatics Vol. 17 no 6 2001.
\bibitem{WR} S. D. M. White \& M. J. Rees, 
Core condensation in heavy halos - A two-stage theory for galaxy formation and clustering,
Mon. Not. R. Astron. Soc. 183, 341, 1978
\end{thebibliography}
\end{document}